\documentclass[%
reprint,
superscriptaddress,
nofootinbib,
nobibnotes,
 amsmath,amssymb,
 aps,
floatfix,
prl,
]{revtex4-2}
\usepackage{graphicx}
\usepackage{dcolumn}
\usepackage{bm}
\usepackage{hyperref}

\usepackage[capitalise]{cleveref}
\usepackage{upgreek}

\begin{document}

\preprint{APS/123-QED}

\title{Suppression and Control of Bipolar Powder Charging by Turbulence}

\author{Simon Jantač}
\email{simon.jantac@ptb.de}
\affiliation{Physikalisch-Technische Bundesanstalt (PTB), Braunschweig, Germany}
\author{Holger Grosshans}%
\affiliation{Physikalisch-Technische Bundesanstalt (PTB), Braunschweig, Germany}%
\affiliation{Otto von Guericke University of Magdeburg, Institute of Apparatus and Environmental Technology, Magdeburg, Germany}

\begin{abstract}
Current models predict particles of the same material but different sizes to charge bipolar upon contacts;
the resulting charge peaks endanger process safety.
However, we found wall-bounded turbulence to suppress the powder's electrostatic charging.
Aerodynamic forces skew the collision frequency and narrow the charge distribution's bandwidth. 
Bipolar charging reduces, especially in moderately polydisperse systems of a low Stokes number. 
Not the smallest but mid-sized particles charge most negatively. 
Moreover, turbulence separates charge, producing pockets of high electric potential in low-vorticity regions.
\end{abstract}

\maketitle

In pneumatic conveyors~\citep{Grosshans2022}, fluidized beds~\citep{Jari2018}, or the atmosphere~\citep{Forward2009,James2008}, particles charge electrically during collisions driven by wall-bounded turbulence.
The particles charge bipolar even if they are of the same material and, thus, lack a work function related charge transfer.
Their polarity depends on their size;
large particles charge positively, and small ones negatively~\citep{Waitukaitis2014, Forward2009phys}.

Conventional Faraday pails measure the sum of all particle charges, they cannot detect charge peaks of both polarities.
Thus, bipolar charge peaks, and their possible discharges, are an immanent hidden threat to powder process safety.

In vacuum and quiescent fluids, simulation models explained the relationship between the particles' charge and size distributions~\citep{KONOPKA2017150,LIU2020199,Xie13};
highly polydisperse systems charge during inter-particle collisions strongly bipolar and slightly polydisperse ones weakly bipolar.
However, the effect of wall-bounded turbulence, which usually drives inter-particle collisions, is unknown so far.

The surface-state theory and mosaic-models specifically aim to predict the charging of insulators with identical, homogeneous surface composition~\citep{Lowell1,Lowell2}.
In their initial formulation~\citep{Lacks2008}, high-energy state electrons can relax into vacant low-energy states.
The high electrical resistivity prohibits high-energy electrons from moving to the vacant low-energy states on the same surface.
Therefore, if the surface contacts another surface, high-energy electrons relax to vacant states of the other surface located at their contact point.
Bipolar charge distributions develop due to asymmetrical charge transfer between particles of different sizes during a collision.
Thus, the frequency of size-dependent collisions determines the resulting charge distribution.
Mosaic-models generalize this concept by substituting surface-states with ambiguous charge carriers, be they ions, electrons, or material patches \citep{Grosjean2023a}.

To sum up, current theories explain bipolar charging by inter-particle collisions but neglect the influence of turbulence.

The essential steps towards understanding the triboelectrification of powder flows require translating the existing knowledge of particle charging to application, bridging the gap from fundamental to applied physics.
To do so, we developed a multi-physics model that differs from previous works by coupling scale-separated physical disciplines:
from the charge exchange in-between particles of the same material during contact up to the interaction of polydisperse particles with wall-bounded turbulence.

The resulting numerical tool can reveal the macroscopic conditions under which hazardous bipolar powder charging happens, thus, opening a new path for process safety.

\begin{figure*}
\centering
\includegraphics[width=\textwidth]{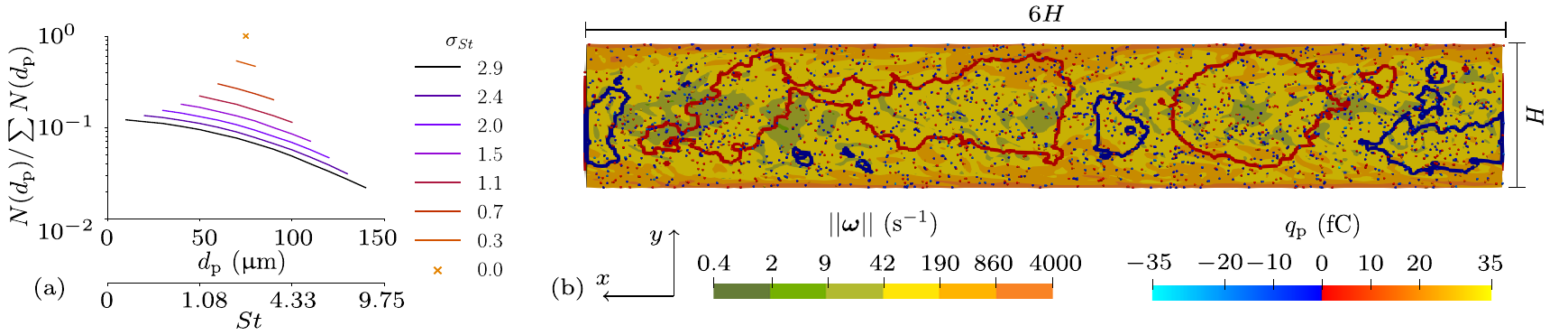}
\caption{a) Normalized particle size distributions tested in this study, from monodisperse ($d_\mathrm{p}=75~\upmu$m, standard deviation of Stokes number of $\sigma_{St}=0$) to most polydisperse ($d_\mathrm{p}=10~\upmu$m--$140~\upmu$m,  $\sigma_{St}=2.9$).
The particles' masses and numbers are the same in all cases, their Stokes numbers differ slightly.
b) 
Snapshot of the simulation domain in flow ($x$) and wall-normal ($y$) direction;
most polydisperse case, the particles obtained their final charge ($t=0.26$~s). 
The vorticity $\Vert \boldsymbol{\omega} \Vert$ (note the logarithmic color map) peaks near the walls.
Pockets of elevated positive electric potential ($>0.3$~V, enclosed by the red isocontour) and negative electric potential ($<-0.3$~V, enclosed by the blue isocontour) form where $\Vert \boldsymbol{\omega} \Vert$ is low. 
}
\label{fig:1}
\end{figure*}

All parts of our tool, the gas phase, particle dynamics, and electrostatic field solvers, were extensively documented and validated earlier~\citep{Gro20d}.
The gas and particle phases are four-way coupled to each other.
That means the particles transfer momentum to the gas, and the particles' trajectories are affected by drag, lift, and collisions with other particles.
Further, the particles interact with the electric field;
charged particles generate the electric field, which then exerts a force on charged particles.

Direct numerical simulations of the Navier-Stokes equations in the Eulerian framework solve the fluid turbulence.
We simulate a wall-bounded turbulent channel flow [see Fig. \ref{fig:1}(b)] of the dimensions $6H \times 2H \times H$ (with $H=4$~cm) in streamwise~($x$), spanwise~($z$), and wall-normal~($y$) direction and a frictional Reynolds number of $Re_\tau=360$.

A grid resolution of $256 \times 144 \times 144$ cells proved to produce grid-independent particle trajectories~\citep{Gro20d}.
The domain is periodic in $x$- and $z$-direction, and the fluid sticks to the walls that confine the domain in $y$-direction.
This generic setup mimics the conditions in which particle-laden flows typically charge;
for example, in pneumatic conveyors, other powder flow devices, or wall-bounded natural flows.

Particles of a number density of $4 \times 10^9$~m$^{-3}$ were seeded randomly into the turbulent gas flow.
The particles are of different sizes but identical material and density, with a particle/fluid density ratio of~1000.

To investigate the effect of polydispersity, we simulated the seven different particle size distributions and one monodisperse distribution depicted in \cref{fig:1}(a). 
At the same time, to isolate the effect of the size distribution, we kept all other parameters constant:
we retained the total solid mass in the system, the number of particles, and the average Stokes number, $\langle St \rangle =\tau_\mathrm{p}/t_0$, the ratio of the particle relaxation time [$\tau_\mathrm{p}=\rho_\mathrm{p} d^2_{32}/(18 \mu)$, where $\rho_\mathrm{p}$ is the particle density, $d_{32}$ the Sauter mean diameter and $\mu$ the fluid's dynamic viscosity] to the flow time-scale.
We defined the flow time-scale as $t_0=H/u_0$, with $u_0$ being the fluid's centerline velocity.
Since the particles' relaxation times scale with their diameters ($\tau_\mathrm{p} \propto d_\mathrm{p}^2$), the standard deviation of the particles' Stokes number distribution, $\sigma_{St}$, quantifies the width of the size distribution.
In addition, we simulated four powders of equal polydispersity (with a Stokes number standard deviation of $\sigma_{St}=2.9$), and varied the average Stokes number from 2 to 39.

The evolution of the particles is tracked in the Lagrangian framework.
Due to their turbopheretic drift, the large particles migrate towards regions of lesser turbulent intensity and small particles to the walls. 
After seeding, the simulations of uncharged particles proceeded until the local average particle concentrations converged.
Then, we activated the below-described charging model.
Thus, all charging simulations presented in this Letter started with fully-developed gas and particle flow fields.

A hybrid method models the interaction of charged particles and the electric field.
Therein, the electric field, calculated by Poisson's equation, exerts far-field forces.
For high accuracy, Coulomb's law directly computes the electrostatic interaction of particles with their close-by neighbors.
The electric potential at the walls equals zero.

The charge of particle $i$ after $N$ collisions with other particles yields $q_{\mathrm{p},i}=\sum_{n=1}^N \delta q_{\mathrm{p},i}(n)$, where $\delta q_{\mathrm{p},i}(n)$ is the charge transferred to the particle during the $n$-th impact.
Thus, we neglect a possible charge transfer to the walls and focus on bipolar charge distributions due to collisions in-between particles.
The collisional charge exchange is computed by the surface-state/mosaic-model.
As mentioned above, according to this model, high-energy state electrons can relax into vacant low-energy states.
More specifically, we followed \citet{KONOPKA2017150} who, instead of describing the transfer of electrons or ions, generalized the model to transferable charged species (TCS) irrespective of their identity.
Under these assumptions, the charge
\begin{equation}
\delta q_{\mathrm{p},i} = \epsilon (c_{\mathrm{s},j}-c_{\mathrm{s},i})A_\mathrm{c,max}^{i,j} 
\end{equation}
transfers from particle $j$ to $i$ during their contact.
In the above equation, $\epsilon$ denotes the charge of one TCS, which is a multiple of an elementary charge.
We assume that the TCS is either an electron or anion with a charge number of -1 (i.e., adsorbed $\mathrm{OH}^-$ \citep{Grosjean2023}, ionomer \citep{Diaz1993}, etc.)
The TCS surface  density on particle $i$ and $j$ before their contact is $c_{\mathrm{s},i}$ and $c_{\mathrm{s},j}$, respectively.
The maximal contact area during the collision, $A_\mathrm{c,max}^{i,j}$, was calculated through the Hertzian theory  \citep{KOREVAAR2014144}.

Reflecting the scarcity of TCS on insulating surfaces, we defined an initial value of $c_\mathrm{s}(0)=10~\upmu$m$^{-2}$ \citep{Lacks2019}.
According to the surface state theory, TCS relax into a stable low-energy state.
However, only TCS in high energy states can exchange during collisions; 
thus, each TCS transfers only once.
In other words, $c_{\mathrm{s},i}$ steadily decreases while $q_{\mathrm{p},i}$ converges to its saturation as the particles undergo collisions.

A flow close to saturation, which means most TCS are transferred, is depicted in Fig.~\ref{fig:1}(b).
Nevertheless, most particles remain airborne instead of adhering to the grounded walls or agglomerating with other particles of opposite polarity.
Thus, all through our simulations, the fluid affects the dynamics of the particles, despite having reached their highest charge.

\begin{figure*}
\centering
\includegraphics[width=\textwidth]{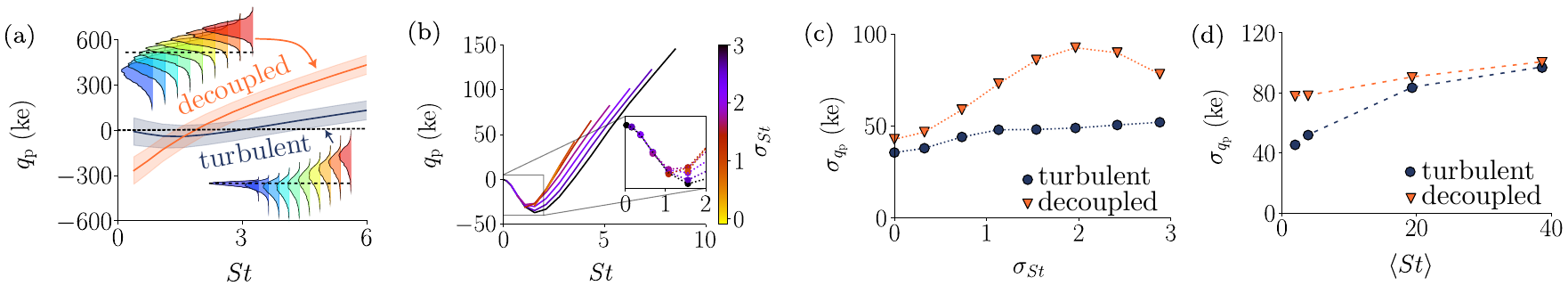}
\caption{a) Resulting charge distribution of particles in turbulence compared to particles that are decoupled from turbulent forces for $\sigma_{St}=2.0$.
The insets give the charge distributions per particle size from $d_\mathrm{p}=30~\upmu$m (dark blue) to $120~\upmu$m (red).
Turbulence suppresses small particles' collisions, thus, skewing the charge distribution.
Particles of  $St \sim 1.5$ obtain the most negative charge.
b) Charge distributions of all polydisperse powders in turbulence when $c_\mathrm{s}= c_\mathrm{s}(0)/e$. 
For $St <1$, the particle charge is independent of polydispersity.
c) Bipolar charging (i.e., standard deviation of the resulting charge distribution) vs.~polydispersity (i.e., standard deviation of the Stokes number) for $\langle St \rangle=3.86$.
d) Bipolar charging vs.~average Stokes number for $\sigma_{St}=2.9$.
Turbulence suppresses bipolar charging the most for a moderately polydisperse powder and small average Stokes numbers. 
}
\label{fig:2}
\end{figure*}

Figure~\ref{fig:2}(a) details the resulting charge distribution when turbulence suppresses bipolar charging the most, namely for $\langle St \rangle =$~3.86 and $\sigma_{St}=1.1$.
The extent of charge suppression is highlighted by comparing the distribution to the charge distribution of the particles whose trajectories are unaffected by turbulent forces, which means they move as in a vacuum.
Even though being decoupled from turbulence, these particles' size distributions are characterized by the corresponding Stokes numbers from the turbulent (coupled) cases.

Decoupling the particles from turbulence leads to a wide bipolar distribution, with the smallest particles carrying the highest negative and the largest particles having the highest positive charge (for $\epsilon=-\mathrm{e}$).
Overall, these observations agree with the state-of-the-art, assuming the particles move unaffected by turbulent forces  \citep{Lee2015,Forward2009phys,Forward2009}.
Affected by turbulence, the particles charge bipolar as well but to a much lesser extent.
Moreover, in turbulence, the charge distribution changes its shape.
The smallest particles carry only a minuscule charge, whereas the particles of $St \sim 1.5$ obtain the highest negative charge.
The largest particles still carry the highest positive charge even though only a fraction of the charge of the same-sized particles that are decoupled from turbulence.
Thus, turbulent forces change bipolar charging not only quantitatively but qualitatively.

For $\langle St \rangle =3.86$, Fig~\ref{fig:2}(b) compares the charge distributions of powders of different polydispersity in turbulence.  
The available TCS decay exponentially while the particles asymptotically reach their final charge.
Therefore, we decided to compare the simulations once the TCS density dropped to $c_\mathrm{s}(0)/e$ ($e$ is Euler's number).
The charge distribution of the particles of a Stokes number below unity is independent of the width of the size distribution. 
The motion of these particles is driven by turbulence, not by their inertia.

Contrary, the charge of the particles of a Stokes number above unity in Fig~\ref{fig:2}(b) depend on the width of the size distribution, namely, the wider the size distribution the more charge they transfer.
The trajectories of these particles are driven by inertia.
Therefore, for $\sigma_{St}<1.5$, where even the smallest particles are of $St > 1$ (cf., \cref{fig:1}), $\sigma_{q_\mathrm{p}}$ increases with $\sigma_{St}$ similarly being affected or decoupled from turbulence, cf., Fig.~\ref{fig:2}(c).

Figures~\ref{fig:2}(c) and~(d) summarize the results of all our simulations.
The standard deviation $\sigma_{q_\mathrm{p}}$ quantifies, analogous to $\sigma_{St}$, the width of the resulting charge distribution.
Since, in sum, all particles in the system are always electrically neutral, $\sigma_{q_\mathrm{p}}$ expresses the extent of bipolar charging.

As elaborated above, by neglecting turbulent dispersion so far, theoretical models predicted large particles to charge positively and small particles to charge negatively;
the broader the size distribution, the broader the resulting charge distribution.
We recovered these previous findings using our model by omitting aerodynamic forces [orange symbols in Fig.~\ref{fig:2}(c)]. 

Even monodisperse particles charge slightly bipolar, namely $\sigma_{q_\mathrm{p}}=42$~ke.
Thus, this amount of bipolar charge is not related to the particle size but to the random collision sequence.
Particles that collide more frequently than the average become net TCS acceptors, and particles that collide less frequently become net donors.
Those particles that collided more often charge in average negative, whereas those particles that collided less often charge in average positive.

\begin{figure*}
\centering
\includegraphics[width=\textwidth]{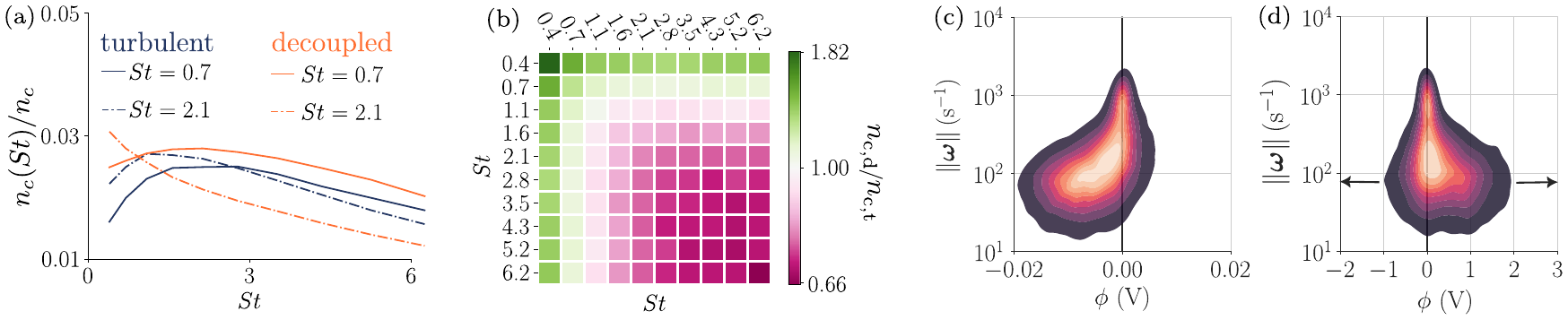}
\caption{
a) 
Comparison of collision frequencies between particles of different sizes in turbulence ($n_\mathrm{c,t}$) and decoupled from turbulence ($n_\mathrm{c,d}$) for $\langle St \rangle =$~3.86 and $\sigma_{St}=2.0$.
Turbulence attenuates the collisions of small particles ($St=0.7$) with all other particles. 
The preferred collision partner of medium-sized particles ($St=2.1$) shifts from $St=0.4$ to $St=1.1$.
b)
When decoupled from turbulence, particles of $St < 2.0$ collide with any other particle more often than in turbulence.
Particles of $St \geq 2.0$ collide in turbulence more often with other particles of $St \geq 2.0$.
Correlation of the electrical potential with the fluid's vorticity for $\langle St \rangle =$~3.86 and  $\sigma_{St}=2.0$  at c) the initial stage ($t=0.2$~s) and d) close to saturation ($t=2.6$~s).
In flow regions of high vorticity, the potential is low.
High potential areas evolve as the charging proceeds in regions of low vorticity (indicated by the arrows). 
 Note the different scaling of the $x$-axes of both figures.
}
\label{fig:3}
\end{figure*}

For polydisperse particles, bipolar charging increases with the width of the size distribution up to the point where the geometrical properties of the smallest and largest particles limit the charge transfer \citep{YU2017113}.
For $\sigma_{St}=2.0$, the standard deviation of the charge distribution reaches its maximal value of $\sigma_{q_\mathrm{p}}=92$~ke.

However, the charge distributions dramatically change if the surrounding turbulence affects the particle trajectories [blue symbols in Figs.~\ref{fig:2}(c) and~(d)].
For monodisperse particles, $\sigma_{q_\mathrm{p}}$ remains low because turbulence affects all particles in the same way.
But for polydisperse systems, turbulence suppresses bipolar charging.
Even though the width of the charge distribution increases with the width of the size distribution, turbulence reduces the gradient by one order of magnitude.
For the highest $\sigma_{q_\mathrm{p}}$, the particles charge only slightly more than the random bipolar charging of monodisperse particles.
At least up to $\sigma_{St}=2.9$, size-dependent charging increases bipolarity only by less than 50\% compared to the charging of monodisperse particles.

Because many particles contained in the size distributions with $\sigma_{St} \ge 1.5$ have a Stokes number less than unity, where their charge is independent of the total width of the size distribution [cf., Fig.~\ref{fig:1}(a) and Fig.~\ref{fig:2}(b)], the increase of $\sigma_{q_\mathrm{p}}$ for $\sigma_{St} \ge 1.5$ [ Fig.~\ref{fig:2}(c)] is slow, nearly plateau-like.   

Next, we checked whether this finding holds for other average Stokes numbers.
To do so, we started simulations keeping a constant polydispersity of $\sigma_{St}=2.9$ but changing the particles' material density to obtain different values for $\langle St \rangle$.
Figure~\ref{fig:2}(d) confirms that turbulence suppresses bipolar charging for low average Stokes numbers. 
For the lowest given Stokes numbers, the width of the resulting charge distribution reduces nearly by half compared to particles decoupled from turbulence.

Nevertheless, the difference between the turbulence-coupled and -decoupled cases becomes smaller for an increasing average Stokes number.
A low Stokes number means the particles exchange little kinetic energy during collisions.
The Hertzian contact area reduces approximately proportional to the kinetic energy;
consequently, the particles exchange little charge.
If less charge is exchanged per collision, depleting the available TCS requires more collisions, which means the charge distributes homogeneously on the particles.
Therefore, unaffected by turbulent forces, $\sigma_{St}$ decreases with decreasing $\langle St \rangle$ [see the orange symbols in Fig.~\ref{fig:2}(d)].
For $\langle St \rangle \to \infty$, inertia exceeds drag, the particles decouple from the flow, and the effect of turbulence diminishes.
In other words, turbulence suppresses bipolar charging for $\langle St \rangle \ll \infty$.

As mentioned above, the suppression of bipolar charging by turbulence relates directly to the inter-particle collision frequencies, see \cref{fig:3} for $\langle St \rangle=3.86$ and $\sigma_{St}=2.0$. 
Decoupled from turbulence, particles of all sizes collide with each other.
Their collision frequencies depend only on their collisional cross-section, which means large particles collide more often than small ones.
When modulated by turbulence, particles get separated according to their size.
In other words, polydispersity, which actually causes bipolarity, attenuates locally.
Then, collisions tend to take place between particles of a similar size.
This leads to a decrease in the collision frequency of the small particles [$St=0.7$ in Fig.~\ref{fig:3}(a)], which means that the charging events occur less frequently than for larger particles. 
Large particles [$St=2.1$ in Fig.~\ref{fig:3}a] collide in turbulence more with other particles of a similar size.

The critical particle diameter is 60~$\upmu$m to 70~$\upmu$m, corresponding to $St=$ 2.1 to 2.8.
Above this value, the collision frequency and charge distribution are monotonous for the Stokes number. 
This is the reason for the suppression of bipolar charging in turbulence because bipolar charging relies on the collisions between different-sized particles.
Since the rate of charge separation is roughly proportional to the diameter difference of the two collision partners, the lower collision frequency of the smallest and largest particles reduces the width of the charge distribution.

To sum up the above discussion, due to turbulence, the preferred location and polarity of particles depend on their size.
Even though being overall electrically neutral, the spatial separation of equally polar particles leads to charged flow regions.
The formation of such regions is depicted in terms of the electric potential by the isocontours in Fig.~\ref{fig:1}(b) and the correlation with the fluid's vorticity in Figs.~\ref{fig:3}(c) and~(d).
The electric potential, because of its relation to the breakdown potential, associates with the explosion hazard of a flow.

During the initial stage of charging, when $c_\mathrm{s}$ is still almost uniform, a negative potential forms.  
This negative potential forms in flow regions where the vorticity is as low as ($\Vert{\bm \omega}\Vert \approx 10^2$~s$^{-1}$), see Fig.~\ref{fig:3}(c).
During the later stage of charging, when the TCS start to deplete, in addition to the negative potential a more pronounced positive potential appears, see Fig.~\ref{fig:3}(d).
From that on, the shape of the potential-vorticity correlation remains stable.
But as particles acquire more charge from collisions, the potential widens in the positive and negative directions, as indicated by the arrows in Fig.~\ref{fig:3}(d).
The areas of high absolute potential ($\vert \phi\vert>$ 0.5 V) remain exclusively in flow regions of low vorticity ($\Vert{\bm \omega}\Vert <500$~s$^{-1}$).

These pockets of high potential form in low vorticity regions because negatively-charged medium-sized particles and positively-charged large particles are expelled from areas of high vorticity.
Contrary, the smallest uncharged particles [cf., Fig.~\ref{fig:2}(a)] remain in areas of high vorticity, leading to regions of low absolute potential ($\vert \phi\vert<$ 0.3~V) for $\Vert{\bm \omega}\Vert >10^3$~s$^{-1}$.

In conclusion, a multi-physics approach revealed a new picture of size-dependent bipolar powder charging:
turbulence changes the size-dependent particle collision frequencies so dramatically that the resulting charge distribution depends on the flow.
More specifically, turbulence drastically reduces the charge distribution's bandwidth.
Contrary to powder decoupled from turbulent forces, in turbulence even strongly polydisperse powder obtains only a slight bipolar charge.
Turbulence spatially separates particles according to their size, thus, reducing the local polydispersity and, consequently size-dependent charging.
In particular, turbulence suppresses bipolar charging the most for particles of moderate Stokes numbers.
Furthermore, decoupled from turbulence, the smallest particles collect the highest negative charge, whereas in turbulence the smallest particles remain nearly uncharged.
Instead, medium-sized particles of $St \approx 1$ gain the highest negative charge.
As a result, in an overall electrically neutral flow, pockets of high positive and negative electric potential form in flow regions of low vorticity.
Generally speaking, the bipolar charging of powder flows depends on charging physics but also strongly on the flow conditions.
Conventional experimental techniques cannot resolve bipolar charge peaks of powder flows.
Thus, the presented computational method and results provide essential steps toward solving a critical problem of industrial process safety.

\begin{acknowledgments}
This project has received funding from the European Research Council~(ERC) under the European Union's Horizon 2020 research and innovation program~(grant agreement No.~947606 PowFEct).
\end{acknowledgments}

\bibliography{References}
\end{document}